\documentclass{PoS}

\title{Dark matter and the early Universe}

\ShortTitle{Dark matter and the early Universe}

\author{\speaker{A.~Arbey}$^,$\thanks{Also Institut Universitaire de France, 103 boulevard Saint-Michel, 75005 Paris, France.}$^{~,1,2}$, J.~Ellis$^{2,3,4}$, F.~Mahmoudi$^{\dagger,1,2}$, G.~Robbins$^{1,4,5}$\\
\\
        $^1$Univ. Lyon, Univ. Lyon 1, CNRS/IN2P3, Institut de Physique Nucl\'eaire de Lyon,\\ UMR5822, F-69622 Villeurbanne, France\vspace*{0.2cm}\\
        $^2$Theoretical Physics Department, CERN, 1 Esplanade des Particules, CH-1211 Geneva 23, Switzerland\vspace*{0.2cm}\\
        $^3$Theoretical Particle Physics and Cosmology Group, Department of Physics, King's College London, London WC2R 2LS, United Kingdom\vspace*{0.2cm}\\
        $^4$National Institute of Chemical Physics \& Biophysics, R{}\"avala 10, 10143 Tallinn, Estonia\vspace*{0.2cm}\\
        $^5$Univ. Lyon, Univ. Lyon 1, ENS de Lyon, CNRS, Centre de Recherche Astrophysique de Lyon UMR5574, F-69230 Saint-Genis-Laval, France\\
        \\
        E-mails: \email{alexandre.arbey@ens-lyon.fr}, \email{John.Ellis@cern.ch}, \email{nazila@cern.ch}, \email{glenn.robbins@kbfi.ee}}

\abstract{Big-Bang nucleosynthesis (BBN) represents one of the earliest phenomena that can lead to observational constraints on the early Universe properties.
It is well-known that many important mechanisms and phase transitions occurred before BBN. We discuss the possibility of gaining insight into the primordial Universe through studies of dark matter in cosmology, astroparticle physics and colliders. For this purpose, we assume that dark matter is a thermal relic, and show that combining collider searches with dark matter observables can lead to strong constraints on the period of freeze-out before BBN.}

\FullConference{The 39th International Conference on High Energy Physics (ICHEP2018)\\
		4-11 July, 2018\\
		Seoul, Korea}

\begin{document}

\section{Introduction}

Big-Bang nucleosynthesis (BBN) provides constraints on the properties of the early Universe. Measurements of the nuclear abundances are in agreement with the assumption that the standard cosmological model is correct up to high energies (i.e., that the early Universe evolution is driven by radiation), which limits possible deviations from the standard cosmological model. However, probing cosmology at temperatures higher than tens of MeV is currently impossible. On the other hand, experiments at colliders, and particularly at the LHC, can probe the standard model of particle physics at the TeV scale. In the following, we combine results from colliders and dark matter observations to set constraints on the cosmological properties of the Universe before BBN. More detail on the analysis can be found in Ref.~\cite{Arbey:2018uho}.

\section{Dark matter relic density}

\begin{figure}
\begin{center}
\includegraphics[width=.6\textwidth]{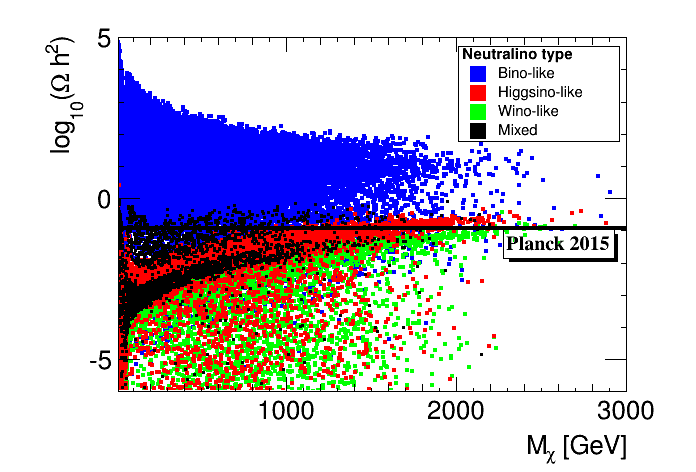}\vspace*{-0.8cm}
\end{center}
\caption{Relic density in the pMSSM as a function of the neutralino mass and type. The horizontal line shows the Planck dark matter density measurement (see Ref.~\cite{Arbey:2017eos}).}
\label{fig1}
\end{figure}

For illustration, we consider the phenomenological MSSM (pMSSM), with the neutralino being the lightest supersymmetry particle, which constitutes a dark matter candidate. To calculate the dark matter relic density, we assume that the neutralino is a thermal relic \cite{Gondolo:1990dk,Edsjo:1997bg}. The neutralino relic density in the pMSSM can take a broad range of values, which can be compared with the Planck measurement of the dark matter density: $\Omega h^2 \sim 0.1$ \cite{Ade:2015xua}. In Fig.~\ref{fig1}, the neutralino relic density is shown for different neutralino types. %
As can be seen, there are still many pMSSM points compatible with all the current cosmological and collider data. The discovery of particles at colliders featuring a relic density compatible with the Planck measurement would constitute a major success for underlying this new physics scenario. If the relic density calculated in the standard cosmological model is too small, one can always consider the possibility that dark matter is made of several components. However, if the calculated relic density is too large, it is generally considered that the scenario is excluded. %
\begin{table}
\begin{center}
\begin{tabular}{|c|c|c|c|c|c|c|c|c|c|c|}
\hline
$M_1$& $M_2$& $M_3$& $\mu$&$M_{A^0}$& $\tan \beta$ & $M_{\tilde{q}_{1,2}}$&$M_{\tilde{q}_{3}}$& $M_{\tilde{l}_{1,2}}$& $M_{\tilde{l}_{3}}$& $A_0$ \\
\hline
-391 &1240 &-1714 &-5739 &4221 & 18.8 &1996 & 4058 & 400 & 1365 & 5372 \\
\hline
\end{tabular}\vspace*{-0.4cm}
\end{center}
\caption{Parameter values of a pMSSM11 benchmark point (in GeV where applicable) that yields too large a relic density when calculated
in the standard cosmological model.\label{point}}
\end{table}%
To illustrate how different properties of the early Universe can change such conclusions, in the next Section we consider the benchmark point with the parameters specified in Table~\ref{point}, which would yield in the standard cosmological model too large a relic density, namely 1.27 (as obtained using SuperIso Relic \cite{Arbey:2009gu}), i.e., one order of magnitude above the Planck value.

\section{Decaying scalar field in the early Universe}

\begin{figure}[b!]
\includegraphics[width=.5\textwidth]{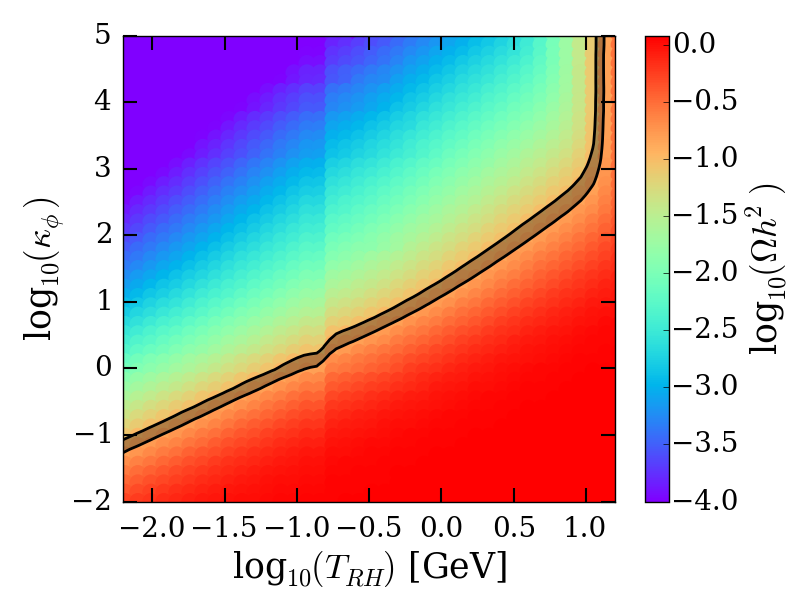}\includegraphics[width=.5\textwidth]{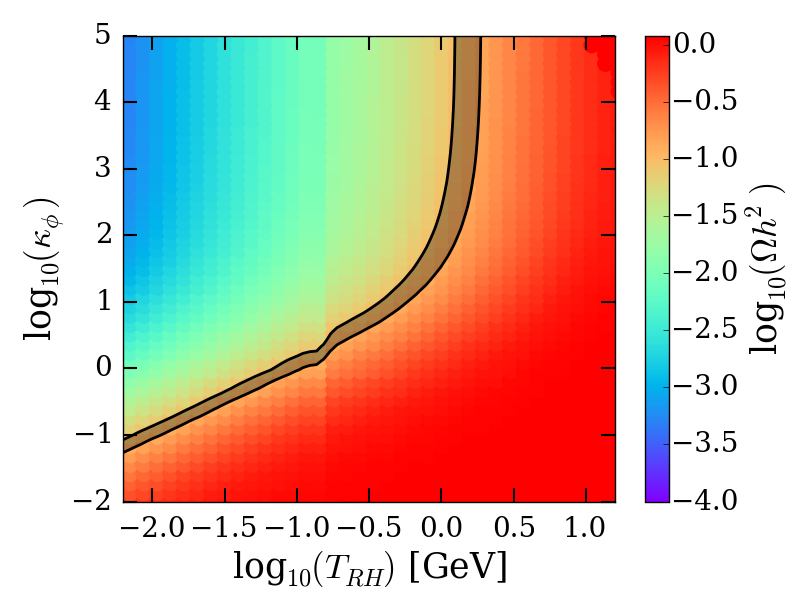}\\
\includegraphics[width=.5\textwidth]{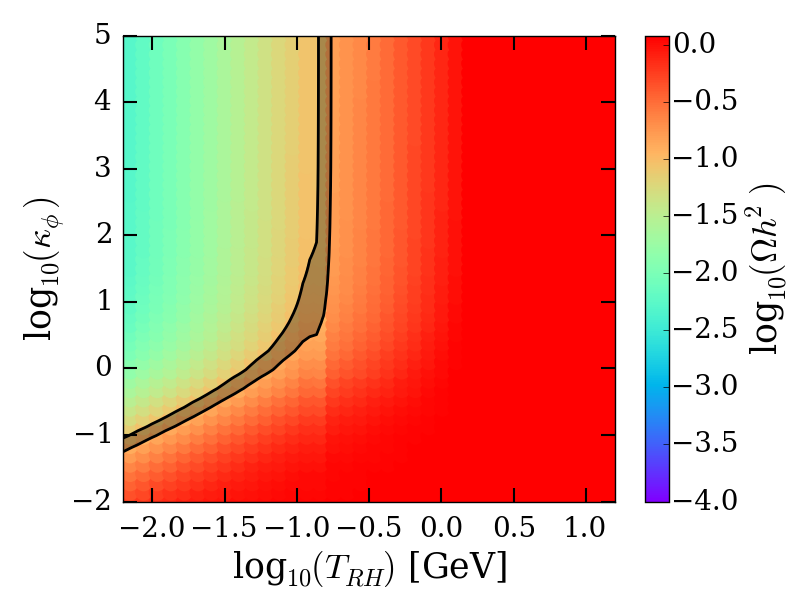}\includegraphics[width=.5\textwidth]{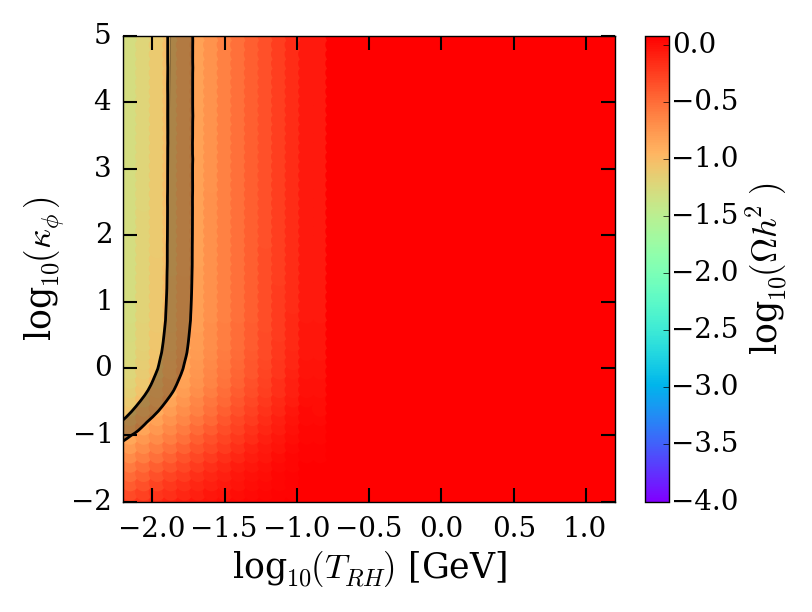}
\caption{Relic density in the parameter plane of the initial scalar field density vs. the reheating temperature, for $\eta_\phi=0$ (upper left), $\eta_\phi=10^{-12}$ (upper right), $\eta_\phi=10^{-11}$ (lower left), $\eta_\phi=10^{-10}$ (lower right). In the dark regions, the relic density is compatible with Planck measurements.}
\label{fig2}
\end{figure}%

In Refs.~\cite{Arbey:2008kv,Arbey:2009gt}, we studied different cosmological scenarios affecting the relic density. In the following, we consider the possibility that a pressureless cosmological scalar field decays into radiation and, with a small branching ratio, into supersymmetric particles around the time of the relic freeze-out. (This scenario is thoroughly described in Refs.\cite{Gelmini:2006pw,Gelmini:2006pq,Arbey:2018uho}.) The three main parameters describing this scenario are: 1) the scalar field density at an initial temperature (40 GeV) normalised to the photon density, $\kappa_\phi$; 2) the reheating temperature $T_{RH}$ at which the scalar field is decayed; 3) the decay branching ratio into relic particles normalised to the scalar field mass in GeV, $\eta_\phi$.

We checked the parameter regions compatible with Big-Bang nucleosynthesis using {\tt AlterBBN}~\cite{Arbey:2011nf}. We then considered how the relic density is affected by the presence of the scalar field in the remaining part of the parameter space. The results are shown in Fig.~\ref{fig2}.
We discuss first the case where the scalar field decay into supersymmetric particles is absent, i.e. $\eta_\phi=0$. For a very small scalar field density, the relic density is unaffected. The same is true for a large reheating density, meaning that the scalar field has already decayed by the freeze-out time. Increasing the initial scalar field density and decreasing the freeze-out temperature result in a decrease in the relic density, induced by the reheating of the early Universe. In this respect, the presence of a decaying scalar field can decrease the relic density by orders of magnitude, while still being compatible with BBN constraints. The case with a decay into supersymmetric particles is similar, with decrease of the relic density for large initial scalar field density and small freeze-out temperature, but the relic density receives a non-negligible contribution from the decay into relics. For this reason, it is also possible to reconcile what would seem to be too small a relic density with the observed dark matter density, because the scalar field decay into relics is not counterbalanced by the reheating.

\section{Conclusion}

The BBN constraints currently constitute the earliest probe of the young Universe. The pre-BBN period is largely unconstrained, and many phenomena could have modified the standard cosmological scenario. In this study, we showed that a decaying cosmological scalar field can modify the relic density by orders of magnitude, showing that the relic density is very sensitive to the properties of the early Universe. Other possible phenomena such as quintessence, phase transitions, moduli, etc., can lead to similar conclusions.

If new particles are discovered in particle physics experiments, and the properties of the underlying scenario are determined, comparing the relic density to the measured dark matter density would allow us to probe the early Universe and, if there is a discrepancy, could lead to the discovery of new cosmological phenomena, as was illustrated here.

\end{document}